\documentclass[namedreferences]{kluwer}

\usepackage{times}
\usepackage{mathptm}

\usepackage{epsfig}

\input epsf
\def\eps@scaling{.95}
\def\epsscale#1{\gdef\eps@scaling{#1}}
\def\plotone#1{\centering \leavevmode
    \epsfxsize=\eps@scaling\columnwidth \epsfbox{#1}}

\def\spose#1{\hbox to 0pt{#1\hss}}
\def\simlt{\mathrel{\spose{\lower 3pt\hbox{$\mathchar"218$}}
     \raise 2.0pt\hbox{$\mathchar"13C$}}}
\def\simgt{\mathrel{\spose{\lower 3pt\hbox{$\mathchar"218$}}
     \raise 2.0pt\hbox{$\mathchar"13E$}}}

\begin{document}

\begin{article}

%
%

\begin{opening}

\title{Rayleigh scattering and laser spot elongation problems at ALFA}

\author{E. \surname{Viard}\email{eviard@eso.org}}
\author{F. \surname{Delplancke}\email{fdelplan@eso.org}}
\author{N. \surname{Hubin}\email{nhubin@eso.org}}
\institute{European Southern Observatory, Garching, Germany}
\author{N. \surname{Ageorges}\email{nancy@epona.physics.ucg.ie}}
\institute{Physics Departement, National University of Ireland, Galway, Ireland}
\author{R. \surname{Davies}\email{davies@mpe.mpg.de}}
\institute{Max-Planck-Institut f\"ur extraterrestrische Physik, Garching, Germany}

\begin{abstract}

This paper describes the qualitative effects of LGS spot 
elongation and Rayleigh scattering on ALFA wavefront 
sensor images.
An analytical model of Rayleigh scattering and a numerical 
model of laser plume generation at the altitude of the 
Na-layer were developed. These models, integrated into a
general AO simulation, provide the sensor sub-aperture
images. It is shown that the centroid measurement accuracy
is affected by these phenomena. The simulation was made 
both for the ALFA system and for the VLT Nasmyth Adaptive Optics 
System (NAOS).

\end{abstract}


\abbreviations{
\abbrev{NGS}{Natural Guide Star};
\abbrev{LGS}{Laser Guide Star};
\abbrev{AO}{Adaptive Optics};
\abbrev{FWHM}{Full Width Half Maximum};
}

\end{opening}

%
%

\section{Introduction} 
\label{sect:intro}  
In LGS-AO systems, many issues have to be studied and tested on real
hardware and the ALFA system in Calar Alto is a remarkable test bench
for the characterization and improvement of LGS-AO operation. 
In Shack-Hartmann systems, it can be observed that the laser
spot shape, as seen from different wavefront sensor sub-apertures,
varies as a function of the launching telescope position relative to
the sensor. The spot shape depends on 2 phenomena : 
\begin{enumerate}
\item the Na-layer thickness : producing a 3-D laser plume of about
10~km which is seen from different angles by the different wavefront
sensor sub-apertures and leads to different laser spot elongations for
each sub-aperture ; 
\item the presence of Rayleigh scattering in the lower atmosphere :
which gives a cone of light below the laser plume itself ; the
Rayleigh cone can be seen by the sensor sub-apertures under certain
circumstances to be studied here. 
\end{enumerate}

\vspace{-1.5cm}
\begin{flushleft}
\begin{figure}[h]
\vspace{1.cm}
\hspace*{0.5cm}\psfig{file=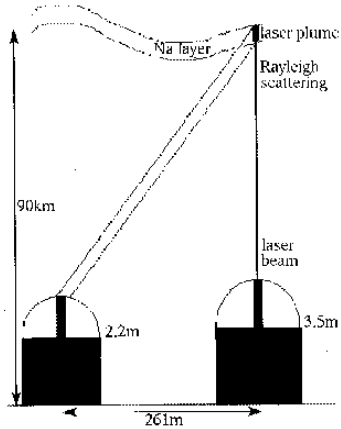,width=4.5cm}
\caption{Calar Alto set-up}
\label{fig:config}
\end{figure}
\end{flushleft}
\vspace{-8.3cm}
\begin{flushright}
\parbox{6cm}{
We have modeled both phenomena independently. Then to calibrate and
verify these t`heoretical and analytical models a common observation
programme was conducted (in August 1998) at Calar Alto. The Na plume
elongation and the Rayleigh cone due to the ALFA laser beam were
observed from the neighboring 2.2~m telescope. The experimental
results \cite{delplancke98} were used to constraint the
models, as well as to study the evolving Na layer density profile. The
experimental set-up is shown in figure~\ref{fig:config}. 
} 
\end{flushright}
The models were introduced into a general AO software package, also
developed in the frame of the LGS network, to get the wavefront sensor
sub-aperture images and the corresponding centroid positions in any
experimental conditions. We have computed the Rayleigh scattering and
the Na spot as seen from wavefront sensor sub-apertures, both in the
ALFA case and in the case of a 8-m telescope (typically the Nasmyth
Adaptive Optics System - NAOS- of the ESO Very Large Telescope
project). 

The influence of the LGS shape on the wavefront sensor performances
and on noise propagation in the AO system can be evaluated. The modal
optimization of the AO closed loop can be adapted accordingly for
optimal results, based on noise propagation. This technique must then
be compared with other proposed ways of removing the perspective
elongation effects \cite{beckers92}.\\
Here are presented the modeling principles, their correlations with
observations, and the qualitative results (sections
\ref{sect:Rayleigh} for the Rayleigh scattering, and \ref{sect:sodium}
for the spot elongation) as well as a first analysis of the LGS shape
influence on measurement noise in section \ref{sect:results}.


\section{Rayleigh scattering modeling}
\label{sect:Rayleigh}
\begin{figure}[htb]
\vspace{-0.5cm}
\centerline{
\raisebox{2.cm}{\hspace*{-0.5cm}\psfig{figure=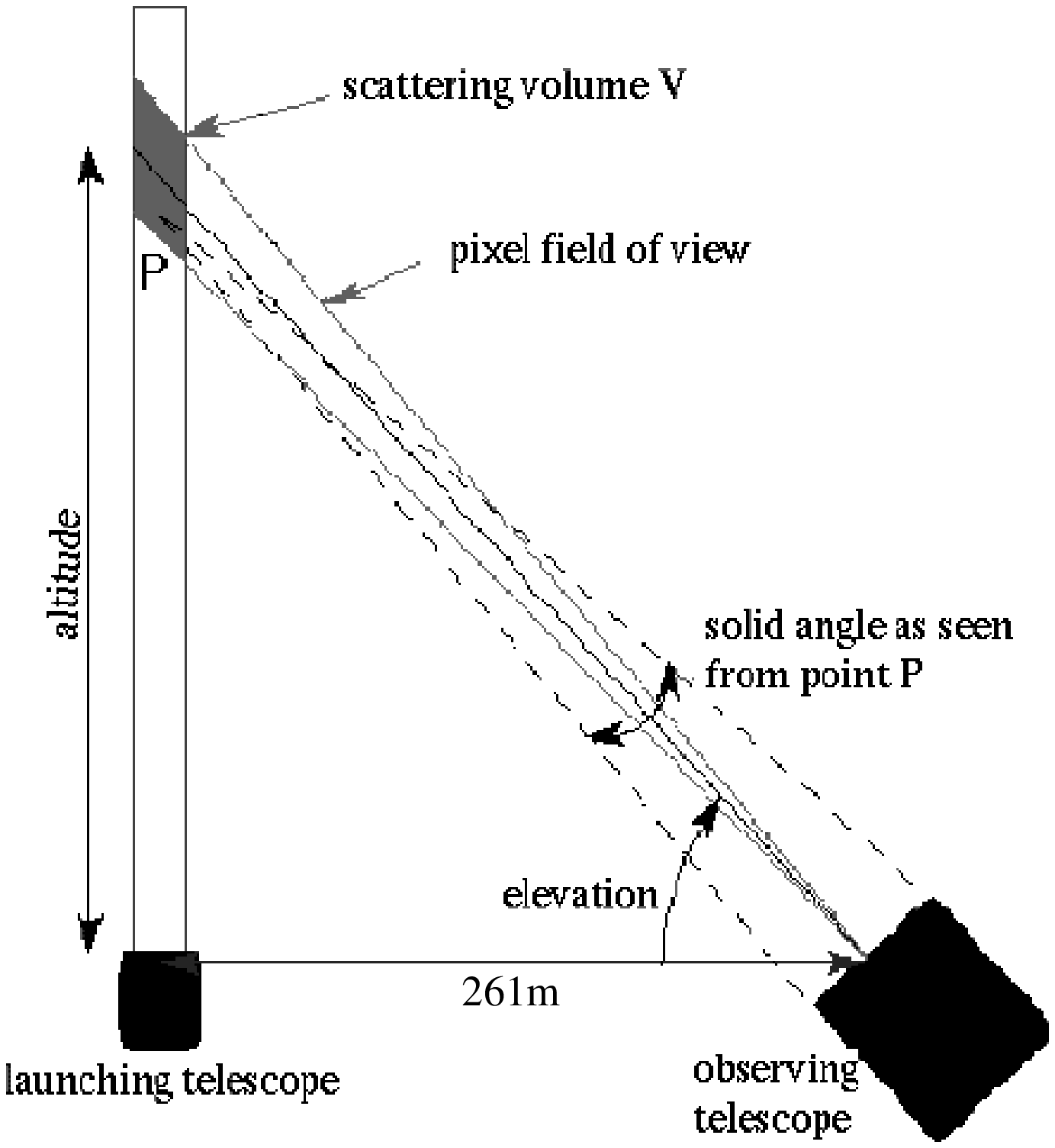,width=6cm}}\qquad
\raisebox{4.5cm}{\hspace*{-1.5cm}\psfig{figure=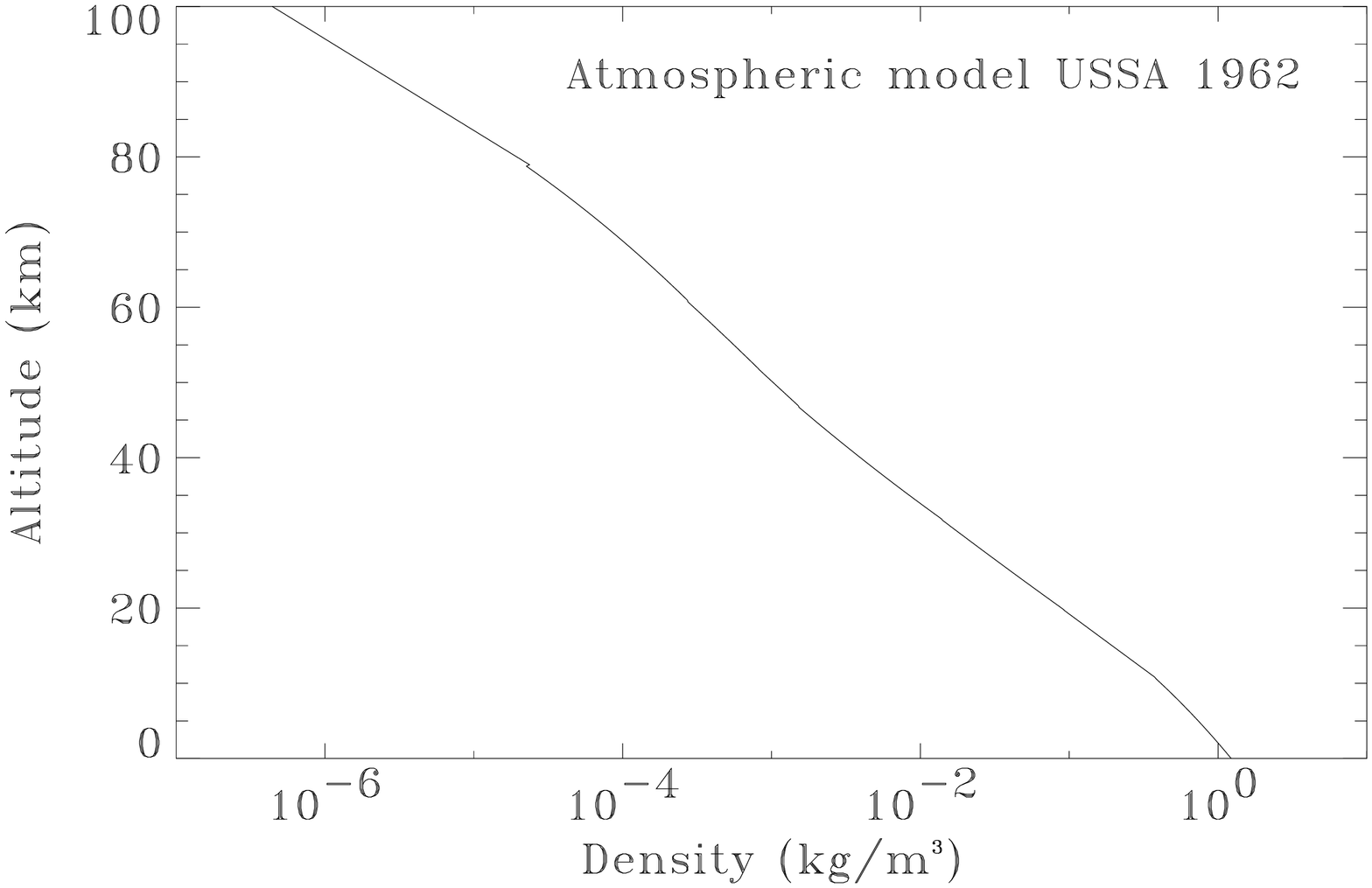,width=5cm,angle=90}}
}
\vspace{-4.5cm}
\caption{Left : Geometry used for the simulation. Right : Atmospheric
density (in $kg/m^{3}$, in log-scale) as a function of the altitude
(in km) as given by the model USSA-1962.} 
\label{fig:geom}
\end{figure}

The Calar Alto experiment was modeled using analytical Rayleigh single
scattering theory and geometrical optics. \\ 
The hypotheses made in the modeling are the following : 
\begin{itemize}
\item the geometry is as shown in Fig.~\ref{fig:geom} with a Na-layer
altitude of 90~km. 
\item the scattering is only due to Rayleigh scatterers (of size $\ll$
than the wavelength), air molecules were used. We did not take the
aerosols into account for 2 reasons : first, their concentration is
highly variable with the period of the year, with the observatory
location, and with the local weather, making the modeling very
difficult. Moreover, as we are interested in the top of the Rayleigh
cone close to the Na-layer, it corresponds to high altitudes where
there are few aerosols. 
\item we considered only single scattering.
\item we neglected the loss of power of the laser beam due to the
scattering when it is going up in the atmosphere (in case of pure
Rayleigh scattering, this loss is lower than $1\%$ on a path length of
100~km). 
\item the beam is supposed to have a constant width all along its
path. Indeed, even if the beam is focalised on the Na-layer, its
diameter does not change significantly due to the turbulence which is
widening the spot. This was verified during Calar Alto run where the
laser plume was observed to have a FWHM diameter of about 1~m (for a
launched beam of 0.25~m). 
\item the beam is assumed wide enough to be resolved by the wavefront
sensor camera on several pixels. Again, it was verified during last
August observations. 
\item the laser launch telescope has no central obscuration.
\item we neglected the influence of the beam polarization because we
are working in backscattering where the scattered light polarization
is identical to the propagating beam one. 
\item the atmospheric density model as a function of the altitude is
the USSA-1962 model \cite{mccartney76a} whose profile is given in
Fig.\ref{fig:geom}. 
\item the Rayleigh scattering analytical formula is the classical one
\cite{vandehulst81} \cite{mccartney76b} :\\ 
\begin{equation}
\frac{dI_1(\alpha)}{d\Omega} = P_0~4~\pi^2~\frac{(n_0-1)^2}{N_0^2~\lambda^4}~(cos(\phi)^2~cos(\alpha)^2+sin(\phi)^2)
\label{equ:scat}
\end{equation}
where $\frac{dI_1(\alpha)}{d\Omega}$ is the intensity of the light
scattered by a single particle, per unit of solid angle $\Omega$ in
the direction $\alpha$ (close to 180$^\circ$ in backscattering), $P_0$
the power of the incident light per unit of surface, $n_0$ the
atmospheric refractive index at sea level (=1.000292), $N_0$ the
number of particles (molecules) per unit of volume at sea level(=$2.5
\times 10^{25}$ particles per cubic meter), $\lambda$ the wavelength,
$\phi$ the angle between the scattering plane and the direction of
polarization of the incident light. 
\end{itemize}

The model predicts correctly the presence of a maximum in intensity
per pixel at a certain elevation depending mainly on the distance
between the launching and observing telescopes, as observed during
Calar Alto observation run from a telescope placed at 261~m from the
launching one. This results from 3 effects :  
\begin{itemize}
\item the decrease in atmospheric density with the altitude, 
\item the increase of the scattering volume (corresponding to one
camera pixel, with the telescope elevation) and 
\item the increase of the distance between that volume and the
observing telescope. 
\end{itemize}

\begin{figure}[h]
\centerline{\psfig{file=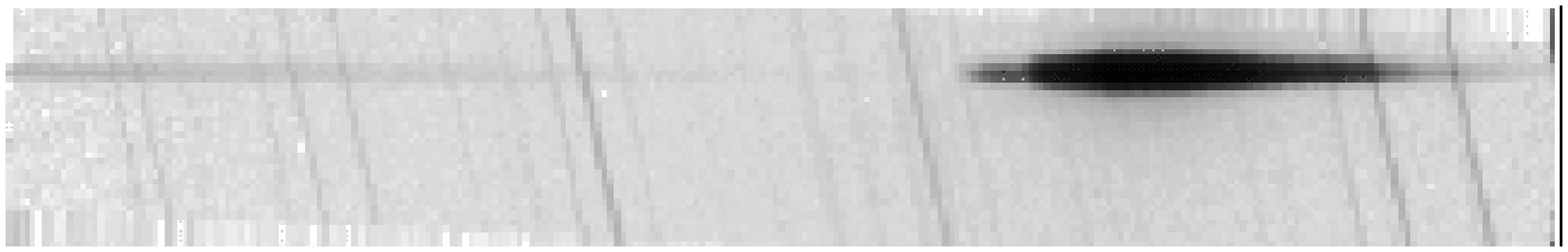,height=11cm,width=1.3cm,angle=270}}
\centerline{\psfig{file=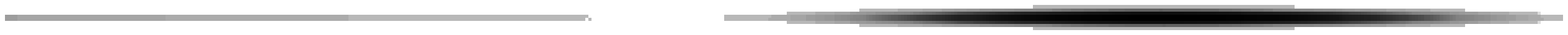,height=11cm,width=1.3cm, angle=270}}
\caption{Na layer plume (right side) and Rayleigh cone top (just seen
at the left side) as seen from the 2.2~m telescope (top) and as
modeled (bottom). The image size is 600x80~$arcsec^2$. Images are in
linear scale} 
\label{fig:top}
\end{figure}

\begin{figure}[htb]
\vspace{-0.3cm}
\begin{center}
\begin{tabular}{cc}
\psfig{figure=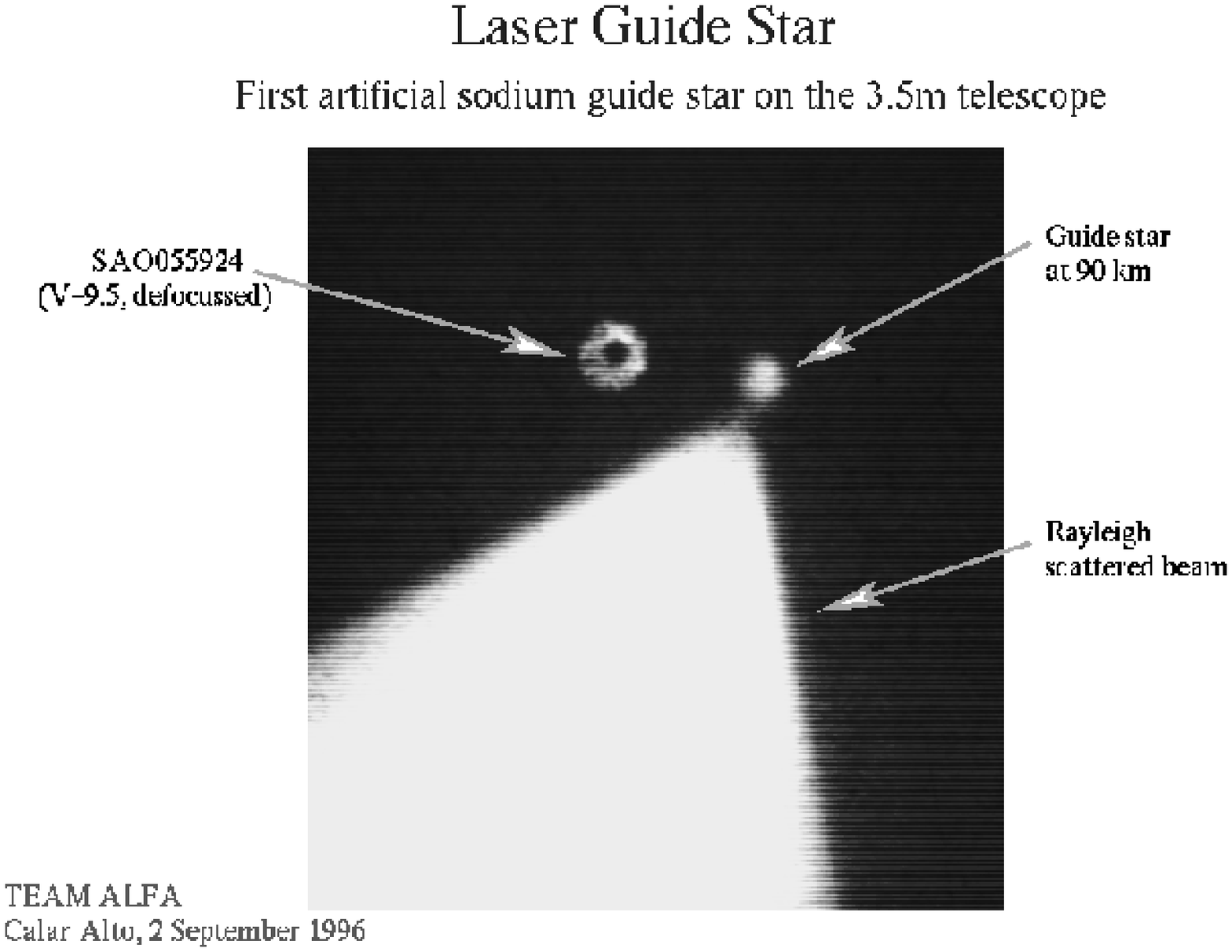,height=5cm,angle=0} &
\raisebox{1.5ex}{\psfig{figure=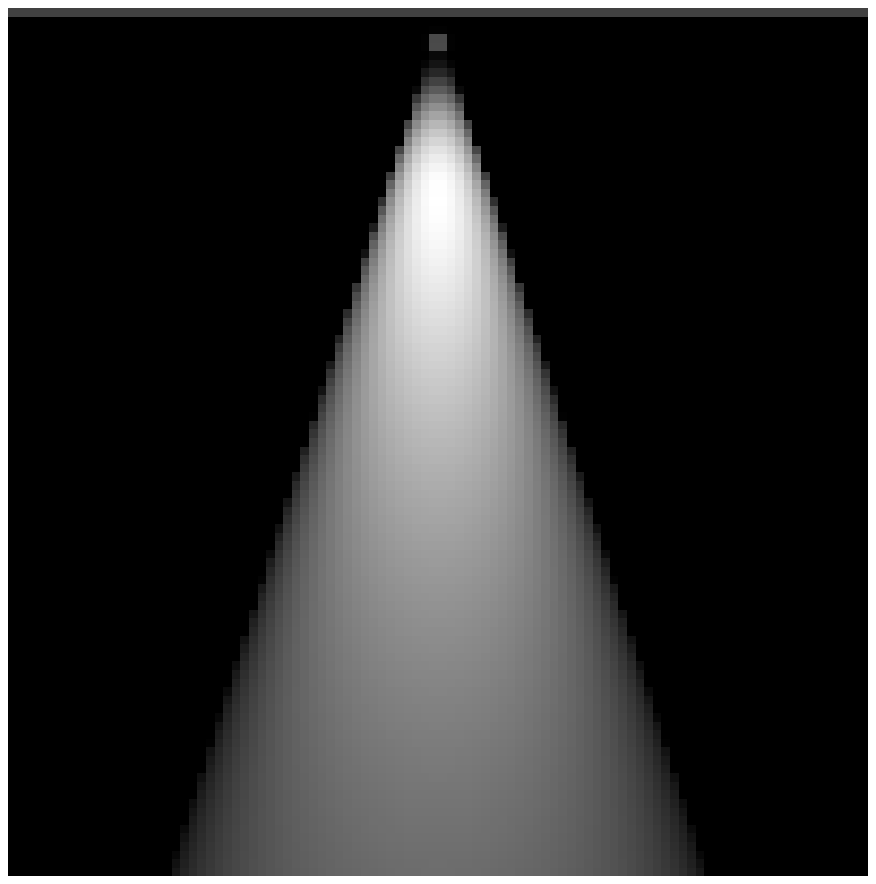,height=4cm,angle=90}}  \\
\end{tabular}
\end{center}
\vspace{-0.55cm}
\caption{Comparison between ALFA off-axis tv-guider image of the LGS
and Rayleigh cone (left) and the image obtained with the simulation
(right) for the same scale and field of view. Distance tv-guider -
launch telescope = 1~m. LGS magnitude = 7. Output laser power =
2.2~W. Pixel size = 0.37~arcsec}
\label{fig:tv_guider}
\vspace{-0.cm}
\end{figure}

The model gives the right shape and the right intensity ratio between
the Na plume (set, in the model, to a cumulated magnitude of 9, as
observed from ALFA \cite{quirrenbach97}) and the top of the Rayleigh
cone as seen in figure~\ref{fig:top}. \\
The simulation gives also results comparable with images taken with
the tv-guider camera of the 3.5~m telescope in Calar Alto
(Fig.~\ref{fig:tv_guider}). The tv-guider camera is much closer to the
laser launch telescope and corresponds more to the close configuration
encountered with the wavefront sensor.

The simulation helps us to predict that the elevation of the apparent
intensity maximum increases when the observing telescope gets closer
to the launching one; the top of the Rayleigh cone approaches the
Na-layer LGS. For close configurations (separation between the
launching telescope and the observing sub-aperture lower than 1~m),
the top of the Rayleigh cone can even apparently {\it touch} the Na
plume.

\section{Sodium layer spot modeling}
\label{sect:sodium}

\begin{figure}[!htb]
\centerline{\psfig{figure=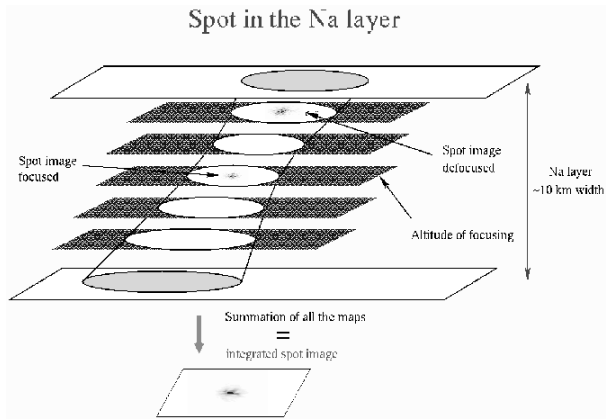,width=5.5cm,angle=270}}
\caption{Principle scheme for the generation of the 3-D spot in the
Na-layer. The sub-layer images are given for a 8-metre launching
telescope in order to show the speckle due to the propagation through
the turbulent atmosphere.}
\label{fig:na_spot}
\end{figure}

\begin{figure}[htb]
\centerline{\psfig{figure=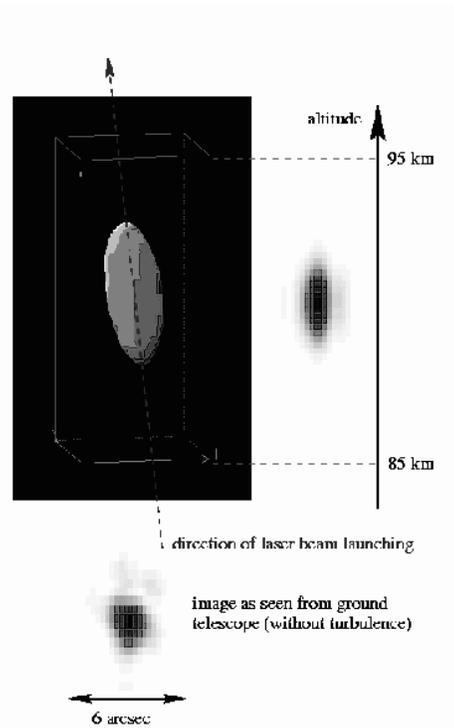,width=6.2cm}}
\caption{Example of a 3-D Na spot obtained with a 0.25~m launched
beam, a 7~km FWHM Na layer at 90~km mean altitude, and a slightly
off-axis launching direction.}
\label{fig:3Dspot}
\end{figure}

The Na layer spot simulation is divided into 3 steps whose parameters are the following :
\begin{itemize}
\item the launch of the laser beam by a telescope of adjustable
diameter (here set to 0.25~m in ALFA case), without central
obscuration. The beam is supposed gaussian, its waist (related to the
full width half max) was chosen here equal to $33~\%$ of the diameter,
its focalization was set on the Na-layer supposed to be at 90~km
altitude, and its power was set to 2~W continuous.
\item the upward conical propagation of the laser beam through a von
K\'arm\'an atmosphere (here with $r_0=20~cm$ and $L_0=20~m$) made of
several layers (here 2 layers at 10~m and 10~km
respectively). Fraunhofer propagation \cite{born59} is assumed here
taking into account the focalization on the Na-layer.
\item the generation of the 3-D spot in the Na-layer by computing the
intensity pattern at different levels in this layer (supposed of FWHM
equal to 7~km centered around an altitude of 90~km), supposing a
gaussian distribution of the Na column density as a function of the
altitude, and not considering saturation (because of the use of a
continuous laser). The resonant backscattered number of photons is
computed following the results obtained by \inlinecite{gardner89}. The
principle scheme of such a simulation is given in
Fig.~\ref{fig:na_spot}. An example of the so obtained 3D spot shape,
as seen from the ground, is given in figure~\ref{fig:3Dspot}.
\end{itemize}


\section{Influence on Shack-Hartmann wavefront sensor performances}
\label{sect:results}

\subsection{Rayleigh cone}
\begin{figure}[!h]
\begin{center}
\begin{tabular}{cc}
\psfig{figure=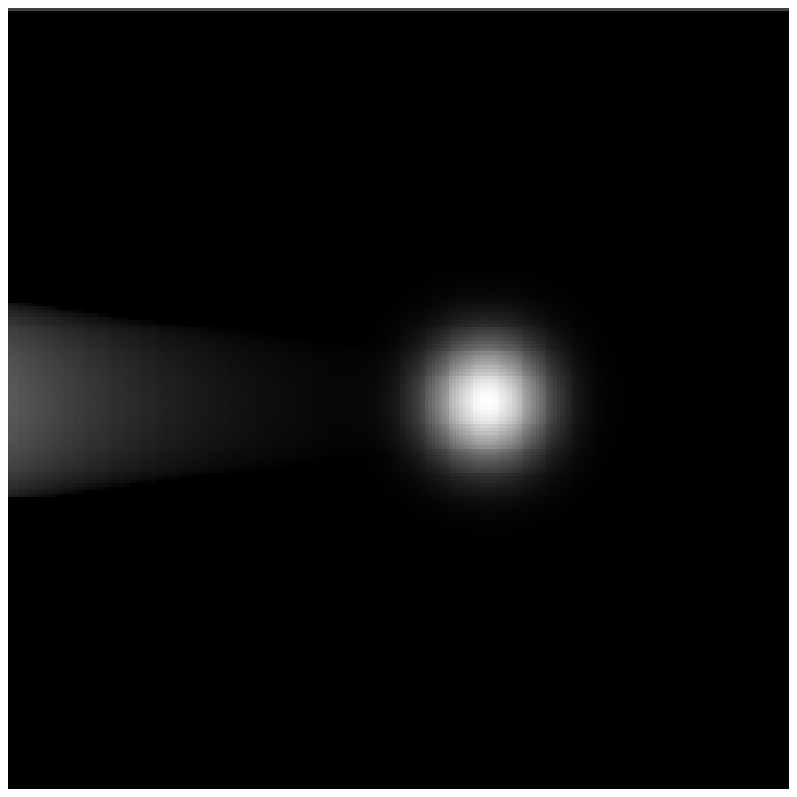,height=2.5cm,angle=0} &
\raisebox{0.ex}{\psfig{figure=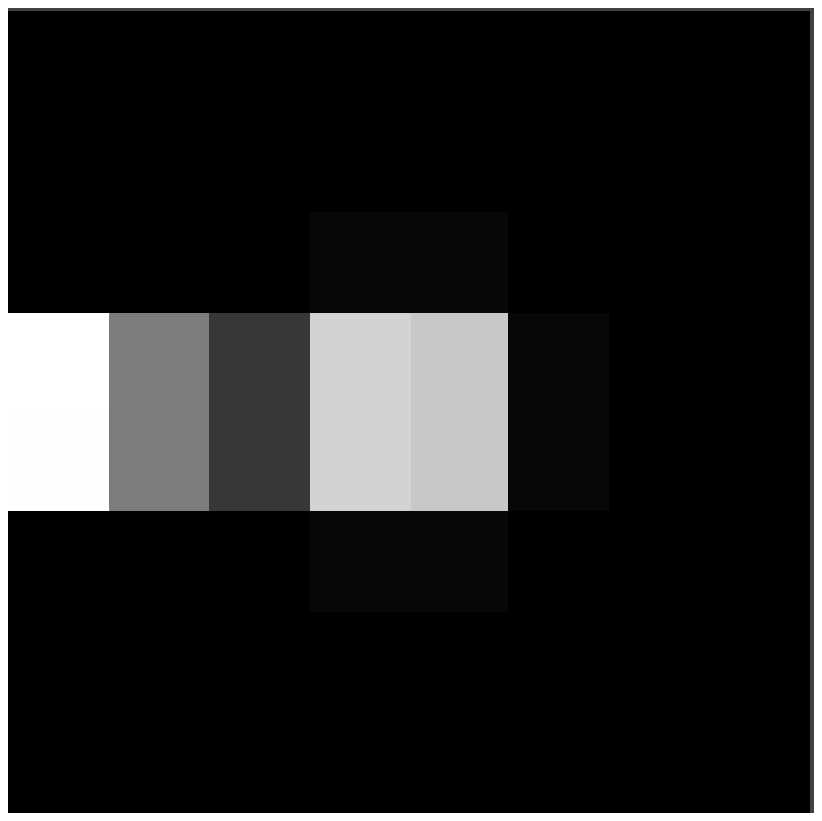,height=2.5cm,angle=0}}  \\
\end{tabular}
\end{center}
\caption{Rayleigh cone and Na spot images in a sub-aperture of
ALFA. The Na spot has a global 9 magnitude, for a 2~W laser power. The
0.7 diameter sub-aperture center is located 1~m from the laser
launching telescope. At left, the pixel size is equal to 0.023~arcsec
while at right, the image is rebinned to the actual ALFA pixel size of
0.75~arcsec. The sub-aperture field of view is in both cases of
6~arcsec. The atmospheric turbulence is not taken into account. Laser
launch telescope diameter = 0.25~m.}
\label{fig:alfa_Rayleigh}
\end{figure}

\begin{figure}[!h]
\centerline{\epsfig{figure=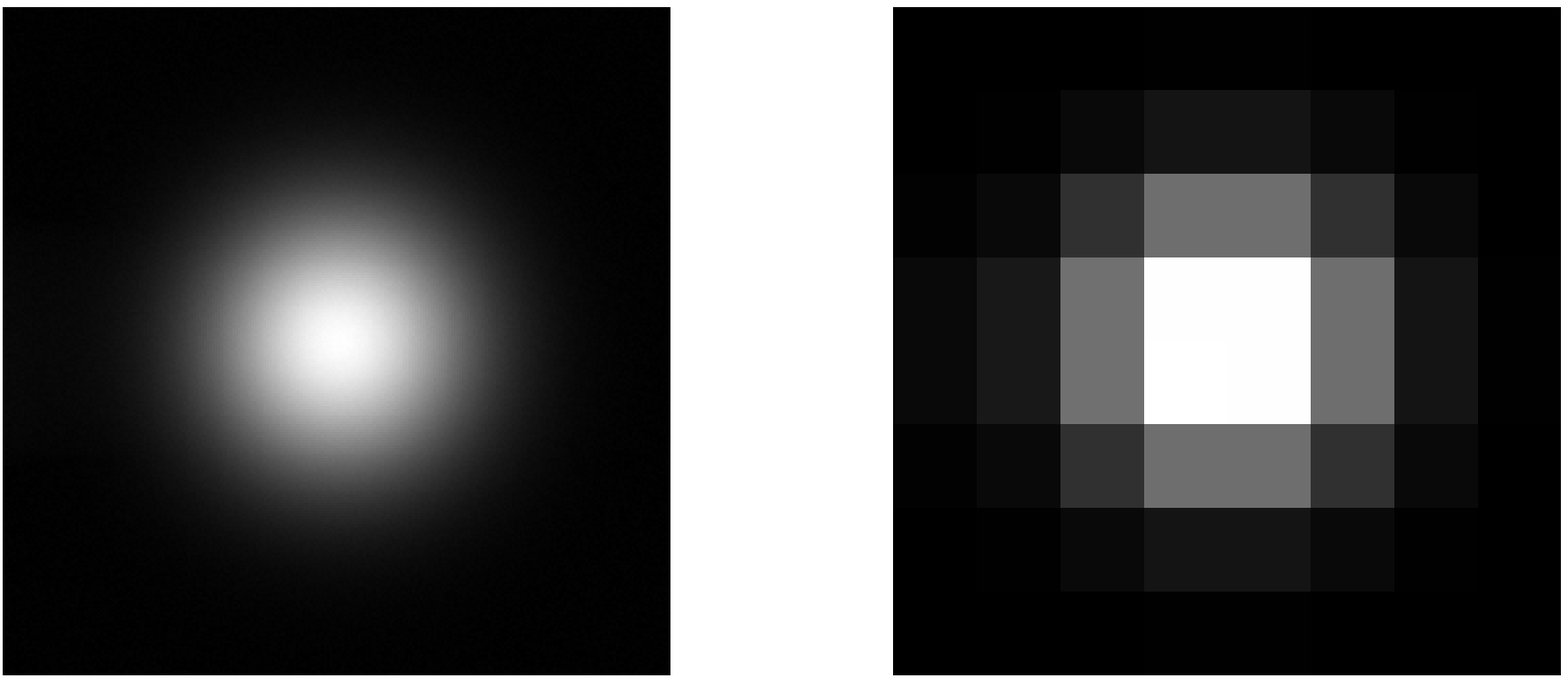, height=2.5cm}}
\caption{Rayleigh cone and Na spot images in a sub-aperture of NAOS
(14x14). The Na spot has a global 9 magnitude, for a 2~W laser
power. The 0.57 diameter sub-aperture center is located 1~m from the
laser launching telescope. At left, the pixel size is equal to
0.009~arcsec while at right, the image is computed for the actual NAOS
pixel size of 0.29~arcsec. The sub-aperture field of view is in both
cases of 2.32~arcsec. The atmospheric turbulence is not taken into
account. Laser launch telescope diameter = 0.25~m.}
\label{fig:naos_Rayleigh}
\end{figure}

As shown by the simulation and by the Calar Alto experiments, the
Rayleigh cone intensity per pixel angularly close to the Na layer
spot, increases when the observing aperture gets closer to the laser
launching telescope. It will be the most important for the closest
wavefront sensor sub-apertures in a LGS AO system. This situation was
modeled both for ALFA (5x5 hexagonal sensor) and NAOS (14x14 square
sensor) systems, considering that the minimum distance between
sub-aperture and launching telescope is equal to 1~m. The results of
such simulation are shown in figures~\ref{fig:alfa_Rayleigh}
and~\ref{fig:naos_Rayleigh}.

In ALFA case, the Rayleigh cone is very important in the sub-aperture,
closest to the laser launch. The total intensity due to the Rayleigh
on the sub-aperture field of view is of the same order of magnitude as
the total Na spot intensity. As the Rayleigh is situated only on one
side of the Na spot, it is inducing a large bias on the centroid
measurement. Fortunately it is not so affecting the other
sub-apertures~: for a laser-sub-aperture distance of 1.7~m, the
Rayleigh total intensity is reduced to about $8~\%$ of the Na spot
total intensity. \\
A solution to remove the Rayleigh scattering cone is to take the
sub-aperture images with a detuned laser so that only the Rayleigh
cone can be seen. This reference image is then subtracted from the
real closed loop image before getting the centroid. It only works when
the laser is launched from the center of the telescope so that the
configuration is centrally symmetric. Indeed, if the laser is launched
sidewise, the pupil rotation on the Shack-Hartmann sensor will induce a
displacement of the laser launch position relative to the
Shack-Hartmann axes. The Rayleigh cone aspect on each sub-aperture will
vary with time and no easy calibration is possible. On the contrary,
when the laser is centrally launched, the pupil rotation does not
induce this displacement and the Rayleigh cone aspect (orientation and
shape) remains the same at any moment (as long as the Rayleigh is
constant). Still, the Rayleigh added noise has to be taken into
account. This parasitic noise is of the order of 5 $e^-$ per pixel per
ms, at his highest point. This is not negligible relative to the other
noise sources : read-out noise (around 6~$e^-$ rms) and sky background
noise.\\
The Rayleigh cone can also be reduced by windowing the sub-aperture
field of view and by temporal filtering in case of pulsed laser. Other
means will be investigated, e.g. using the beam polarization
properties.

On the contrary, Fig.~\ref{fig:naos_Rayleigh} shows that the Rayleigh
cone can be neglected in the case of NAOS. 
This is due to the fact that, in NAOS, the pixel scale is much smaller
than in ALFA (0.29~arcsec versus 0.75~arcsec). The sub-aperture field
of view is thus smaller, providing a kind of spatial
filtering. Moreover, in NAOS, the Rayleigh cone is distributed on a
larger number of pixels than in ALFA while the Na spot is not {\it
diluted} so much. The intensity per pixel due to the Rayleigh
scattering goes down to $4~\%$ of the Na spot intensity per
pixel. This represent a parasitic noise, after Rayleigh cone
substraction, of 0.4~$e^-$ per pixel per ms, which is negligible
relative to the read-out noise. 

\subsection{Spot elongation}
\begin{figure}[htb]
\centerline{\psfig{figure=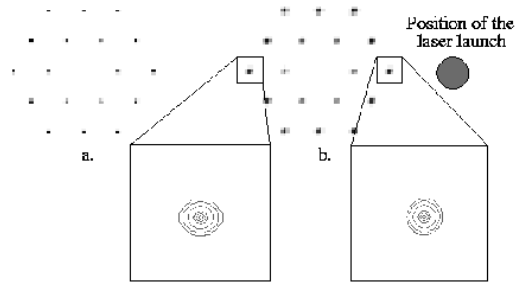,width=7.5cm,angle=0}}
\caption{Right : Na spot elongation on the various sub-apertures of
ALFA system (5x5 hexagonal array). The laser is launched from the
pupil side (grey dot). The closest and farest spots are enlarged as
contour plots in order to show the differential spot elongation. The
global field of view of the contour plots is 3~arcsec. It can be
compared with the natural guide star spot shape (at left).}
\label{fig:alfa_spots}
\end{figure}

\begin{figure}[htb]
\centerline{\psfig{figure=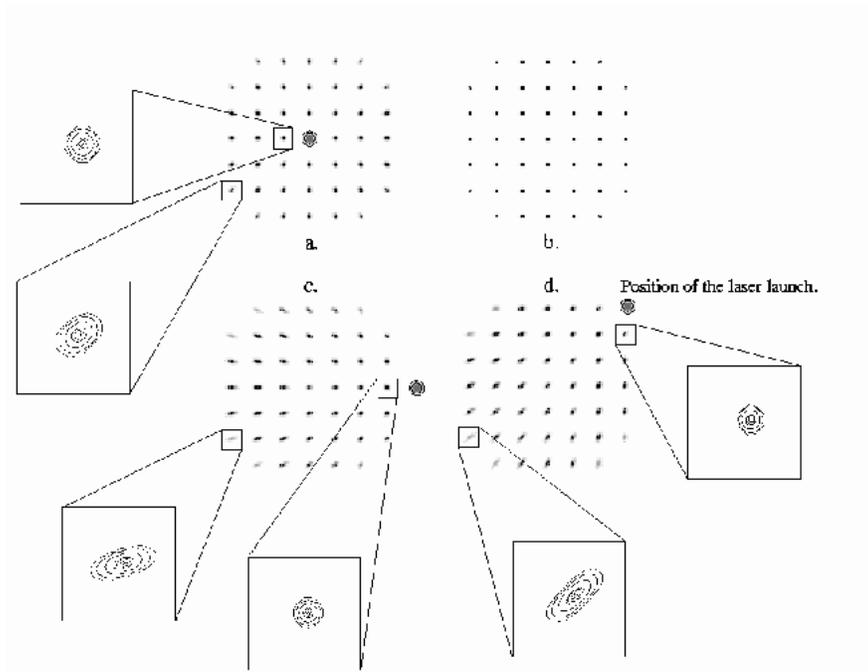,width=9cm,angle=270}}
\caption{Na spot elongation on the various sub-apertures of NAOS
system (8~m telescope, 7x7 square array) for various positions of the
laser launch telescope (grey dot) : a) behind the secondary mirror, c)
at the pupil side, along one of the sensor axes, d) at the pupil side
and at 45$^\circ$ from the axes. It can be compared with the natural
guide star spot shape (b). Some of the spots are enlarged into contour
plots whose global field of view is 3~arcsec.}
\label{fig:naos_spots}
\end{figure}

The 3-dimensional Na spot image, obtained as described the
section~\ref{sect:sodium}, is then used in the wavefront sensor image
computation. \\
Again Fraunhofer and weak fluctuations approximations are used : the
scintillation is neglected and we propagate only phase variations
through the same atmosphere as used in upward propagation, taking the
cone effect into account. Phase at the entrance of the telescope is
obtained and Shack-Hartmann wavefront sensor sub-aperture images are
computed using Fourier optics~: 
\begin{itemize}
\item the 2-D Na spot source maps are obtained by projecting the 3-D
map on each sub-aperture taking the perspective effects into account~;
\item the sensor point-source images are computed by Fourier transform
of the phase on the sub-apertures~; 
\item the convolution of the point-source images with the 2-D Na spot
maps gives us the final sensor images. 
\end{itemize}
This simulation was made in 2 cases : ALFA (3.5~m telescope with a 5x5
hexagonal Shack-Hartmann wavefront sensor) and NAOS (8~m telescope
with a 7x7 square Shack-Hartmann sensor). \\
The 2-D Na spot maps are shown in Fig.~\ref{fig:alfa_spots} for ALFA
and in Fig.~\ref{fig:naos_spots} for NAOS and for different positions
of the laser launch telescope relative to the pupil. The differential
elongation of the spots as a function of the sub-aperture can be
easily seen on the extracted contour plots. It is more noticeable with
the 8~m telescope than with the 3.5~m.

The spot elongation modifies the accuracy of the centroid measurement
along its long axis. As the spot elongation is not the same for each
sensor sub-apertures and as a function of the laser position, the
noise on the centroiding measurement is different and therefore the
noise propagation into the reconstructed modes is affected. This
effect is negligible for ALFA. Indeed the measurement noises (read-out
noise, background noise ...) are very large relative to the LGS signal
and the laser spots are deformed due to atmospheric turbulence. The
very slight spot elongation shown in figure~\ref{fig:alfa_spots}
without noise nor perturbation is so completely dominated by other
ALFA noise sources in real situations. \\
On the contrary, in system like NAOS where the measurement noises are
considerably reduced, the spot elongation will have to be taken into
account and the laser launch telescope position well chosen. Indeed,
in an alt-azimutal telescopes, the pupil rotates on the sensor and if
the launching laser is not located behind the telescope secondary
mirror (central position), the shape of the spots on each sub-aperture
will vary with time. This effect is well shown in
Fig.~\ref{fig:naos_spots} c and d. \\
In LGS AO operation, the differential noise propagation must be
compensated via the reconstruction matrix and via an adapted modal
optimization in the time and mode filtering. If the spot elongation on
one sub-aperture varies with time (i.e. with a lateral launching
telescope), various reconstruction matrix have to be implemented as a
function of the pupil position on the Shack-Hartmann sensor. The
quantitative effects of these phenomenon on the adaptive optics loop
performances have to be evaluated.

\section{Conclusion and Perspectives}
Experiments were conducted in Calar Alto with the 2.2 and 3.5~m
telescopes and with ALFA, in order to study the Rayleigh scattering
induced by the laser beam, and the spot elongation due to the
non-negligible Na-layer thickness.\\
The observations allowed us to develop and calibrate models of these
phenomena. The models have been used to study the impact of Rayleigh
scattering and spot elongation on the Shack-Hartmann wavefront sensor
image and on the centroiding performances. 
The spot elongation was shown to be negligible in ALFA case while it
is more important with an 8~m telescope. On the contrary, the presence
of the top of the Rayleigh scattering cone appeared to be more
noticeable in ALFA case and to induce a non-negligible bias in the
centroiding measurements close to the laser launch telescope.

We are planning to make new observations in Calar Alto to exactly
calibrate the photon returns from the Na spot and from the Rayleigh
cone, together with Lidar measurements of the atmosphere. These data
will be used to certify our developed models. \\
We will also test the polarization method for suppressing the Rayleigh
cone in the sub-aperture images.
Finally, the investigation of noise propagation in the AO loop with
spot elongation will be pursued quantitatively. We plan to propose
strategies to correct for this effect.

%
%

%

\begin{acknowledgements}
This work is being performed in the frame of the `Laser Guide Star for 8-m Class Telescopes''
network of the European Commission {\it Training and Mobility of
Researchers} programme (Contract no. FMRX-CT96-0094). The present
research was a collaboration between
 the Max-Planck-Institut f\"ur
extraterrestrische Physik in Garching, the European Southern
Observatory and the National University of Ireland-Galway.\\ 
Special thanks go to the ALFA team (W. Hackenberg, M. Kasper, Th. Ott
and S. Rabien) and the operators (J. Aceituno and L. Montoya) for
their essential help with the laser.  
\end{acknowledgements}

%
%

%

%
\bibliographystyle{klunamed}
\bibliography{biblio2}

\begin{thebibliography}{}

\bibitem[\protect\citeauthoryear{{Beckers}}{1992}]{beckers92}
{Beckers}, J.: 1992, `Removing perspective elongation effects in Laser Guide
  Stars and their use in the ESO Very Large Telescope'.
\newblock {\em Proceedings of ESO conference on Progress in Telescope and
  Instrumentation Technologies (April 27-30, 1992, Garching, Germany)} pp.
  505--514.

\bibitem[\protect\citeauthoryear{{Born} and {Wolf}}{1959}]{born59}
{Born}, M. and E. {Wolf}: 1959, {\em Principles of Optics}, pp. 381--385.
\newblock Pergamon Press (London).

\bibitem[\protect\citeauthoryear{{Delplancke} et~al.}{1998}]{delplancke98}
{Delplancke}, F., N. {Ageorges}, N. {Hubin}, and C. {O'Sullivan}: 1998, `LGS
  light pollution investigation in Calar-Alto'.
\newblock In: {\em Proceedings of ESO-OSA Conference on Astronomy with Adaptive
  Optics ( April 7-11, 1998, Sonthofen, Germany)}. pp. 501--512.

\bibitem[\protect\citeauthoryear{{Gardner}}{1989}]{gardner89}
{Gardner}, C.: 1989, `Sodium resonance fluorescence Lidar applications in
  atmospheric science and astronomy'.
\newblock In: {\em Proceedings of IEEE}, Vol.~77. pp. 408--418.

\bibitem[\protect\citeauthoryear{{McCartney}}{1976a}]{mccartney76a}
{McCartney}, E.: 1976a, {\em Optics of the Atmosphere - Scattering by Molecules
  and Particles}, pp. 81--86.
\newblock John Wiley \& Sons (New York).

\bibitem[\protect\citeauthoryear{{McCartney}}{1976b}]{mccartney76b}
{McCartney}, E.: 1976b, {\em Optics of the Atmosphere - Scattering by Molecules
  and Particles}, Chapt.~4.
\newblock John Wiley \& Sons (New York).

\bibitem[\protect\citeauthoryear{{Quirrenbach} et~al.}{1997}]{quirrenbach97}
{Quirrenbach}, A., W. {Hackenberg}, H. {Holstenberg}, and N. {Wilnhammer}:
  1997, `The Sodium Laser Guide Star System of ALFA'.
\newblock In: {\em Proceedings SPIE in Adaptive Optics \& Applications}, Vol.
  3126. pp. 35--43.

\bibitem[\protect\citeauthoryear{{van de Hulst}}{1981}]{vandehulst81}
{van de Hulst}, H.: 1981, {\em Light scattering by small particles}, p.~65.
\newblock Dover Publications (New York).

\end{thebibliography}
\end{article}
\end{document}